\documentclass[showpacs,aps,twocolumn]{revtex4/revtex4}
\usepackage{epsfig}
\usepackage{graphicx}
\usepackage{amsmath,amssymb,amsfonts}
\usepackage{array}
\usepackage{url}
\usepackage{hyperref}
\usepackage{multirow}
\usepackage{float}
\usepackage{lineno}
\usepackage{xspace}
\usepackage{ulem}
\usepackage[usenames,dvipsnames]{color}

\begin{document}
\title{Fluidity of the system produced in relativistic pp and heavy-ion collisions:  Hadron resonance gas model approach}
\author{Ronald Scaria$^1$}
\author{Dushmanta Sahu$^1$}
\author{Captain R. Singh$^1$}
\author{Raghunath Sahoo$^{1,2}$\footnote{Corresponding author: $Raghunath.Sahoo@cern.ch$}}
\author{Jan-e Alam$^3$}
\affiliation{$^{1}$Department of Physics, Indian Institute of Technology Indore, Simrol, Indore 453552, India}
\affiliation{$^{2}$CERN, CH 1211, Geneva 23, Switzerland}
\affiliation{$^{3}$Variable Energy Cyclotron Centre, 1/AF, Bidhan Nagar, Kolkata, India}

\begin{abstract}
We have estimated the dimensionless parameters such as 
Reynolds number ($Re$), Knudsen number ($Kn$) and Mach number ($Ma$) for a
multi-hadron system by using the excluded volume hadron resonance gas (EVHRG) model along with 
Hagedorn mass spectrum to include higher resonances in the system.
The size dependence of these parameters indicate that 
the system formed in proton+proton collisions may achieve thermal equilibrium 
making it unsuitable as a benchmark to analyze the properties of the system produced 
in heavy ion collisions
at similar energies. While the magnitude of $Kn$ can be used to study the degree of thermalization
and applicability of inviscid hydrodynamics,
the variations of $Re$ and $Ma$ with temperature ($T$) and baryonic chemical
potential ($\mu_B$) assist to understand the change in the nature of the flow in the system.
Indeed the nature of flow changes from laminar to turbulent as $Re$ increases and the
system is characterized as incompressible for low $Ma (<<1)$ and compressible for larger
$Ma$.  $Ma$  can also be used to understand whether the flow is subsonic or supersonic.
\end{abstract}
\date{\today}
\maketitle 

\section{Introduction}
Relativistic heavy-ion collisions aim to create extreme conditions of temperature ($T$)  and baryonic chemical potential ($\mu_B$)
so as to melt the hadrons to produce a deconfined state of quarks and gluons, called Quark-Gluon Plasma (QGP). 
Although initially it was envisaged that such a state of matter would behave as an ideal gas, 
the analysis of the results from the nuclear collisions at the Relativistic Heavy-Ion Collider (RHIC) and Large Hadron Collider (LHC) energies 
reveal that the created matter behaves like strongly interacting liquid with low shear viscosity ($\eta$) to entropy density ($s$) ratio. 
The study of 
experimental data within the framework of relativistic hydrodynamics 
shows that the $\eta/s$ of QGP is close to $1/4\pi$, which is the AdS/CFT bound obtained in Ref. \cite{Kovtun:2004de}, indicating 
the creation of a ``perfect fluid" in relativistic heavy-ion collisions at the RHIC \cite{Arsene2005,Back2005,Adams2005,Adcox2005}. 
So far, the results from proton+proton (pp) collisions were used as a baseline for 
establishing QCD medium formation in heavy-ion collisions at RHIC and LHC energies.  
However, recent observations of strangeness enhancement, collectivity, long-range particle correlations etc. in pp and p+Pb collisions 
at the LHC energies~\cite{Adam2017, Khachatryan2017} opened up a new domain of studies. These observations advocate for the 
possibility of thermalization and collectivity in the small system  produced  in pp collisions. 
Final state multiplicity dependent event characterization 
and collision centrality selection even in such collisions emerged
out with intense search for QGP-like properties in high-multiplicity pp collisions. Although the formation of
QGP-droplets is yet to be established, a detailed characterization of the system formed in pp collisions at the LHC
through various theoretical models and experimental searches using new observables are underway. 
The study of final state event multiplicity in pp, p+Pb and heavy-ion collisions at the LHC 
energies, has become one of the unique observables to describe the system size and collision 
energy dependence of various observables. In this 
direction, one tries to find a threshold in the final state multiplicity \cite{Velicanu:2011zz,Campanini:2011bj,Thakur:2017kpv,Sahu:2020nbu} after which QGP-like properties reveal and the
corresponding theoretical models to understand a QCD medium formation, including those to study thermalization and  
hydrodynamic models could be employed.

The possibility of thermalization in a system and the applicability of hydrodynamics can be studied using $Kn$. 
The nature of the hydrodynamic flow (laminar or turbulent) can be classified based on 
the magnitude of the  $Re$ and the $Ma$ helps in determining 
whether the fluid is compressible or incompressible. The inequalities, $Ma<<1$ (subsonic) indicate
that the fluid is incompressible and $Ma>>1$ (supersonic) hints that it is compressible.
Fluid with a high value of $Re$ will show turbulent behavior. $Kn$ is the ratio of the mean free 
path of the system to some macroscopic length scale. Therefore, small values of $Kn$ imply a system with 
a small gradient of hydrodynamic quantities (see~\cite{hydrobook} and references therein) and a high degree of thermalization. 
A large value of $Kn$ suggests that the system remains far away from equilibrium.
In such cases fluid dynamics can still be applied with the inclusion of higher-order gradients 
of hydrodynamic quantities which is a very active field of contemporary research
but beyond the scope of present work (see~\cite{hydrobook} for details). 
The $Kn$ was estimated from 
a hydrodynamic model, in which the elliptic flow as a function of 
centrality suggests the value of 
$Kn$ $\simeq$ 0.3 for central collisions and 0.5 for semi-central 
collisions~\cite{Bhalerao:2005mm,Drescher:2007cd}
of Au$+$Au at $\sqrt{s_{\rm NN}}$ = 200 GeV. 
As $Re\sim \frac{inertial force}{viscous force}\sim 1/\eta$, systems with low viscosity are 
likely to manifest turbulent nature because in such cases perturbations in hydrodynamic 
quantities will grow and drive the system towards turbulent motion.
At RHIC energy for a system of size 5 fm and temperature around $T = 300$ MeV, $Re$ varies 
between 10$-$100 for a fluid velocity in the range 0.1$-$1 
(in the unit of $c=1$)~\cite{Deng:2016gyh}. Similarly, $Ma$ is defined as the ratio of the 
flow velocity to the speed of sound in the system. For $Ma$ $<<$ 1, the fluid density is almost 
uniform which suggests that the flow is incompressible. Thus, $Ma$ can help in 
characterizing the compressibility of the fluid. These three parameters can be used together
to achieve the following goals: (i) $Kn$ will indicate the 
thermalization in the system, (ii) $Re$ hints at the nature of the flow (laminar or turbulent)
and finally (iii) $Ma$  indicates whether the fluid is 
compressible ($Ma\gtrsim 1$)  or incompressible  ($Ma<<1$). The $Ma$
for  relativistic fluids (flow velocity close to the velocity of light) will be larger
than 1 indicating that it can not be treated as incompressible.  Therefore, it is expected 
that fluid system produced in the relativistic collision of nuclei will be compressible.
The applicability of hydrodynamics (for $Kn<<1$) will give information about properties 
like compressibility and viscosity of the fluid ~\cite{Ollitrault2008}. 

To understand the thermodynamics of the system produced in ultra-relativistic collisions, 
inputs from lattice quantum chromodynamics (LQCD) have been extremely useful. However, it works 
accurately for very low baryochemical potential ($\mu_B/T\rightarrow 0$). In this context models such as Hadron Resonance Gas (HRG) can be useful to analyze data 
without having restrictions on the value of baryochemical potential. The HRG model has been used successfully to study the thermodynamics 
of the hadronic phase that appears in the course of the evolution of the system produced in ultra-relativistic collisions. In this article, 
we have used the HRG model with excluded volume effect along with a Hagedorn mass spectrum to study the thermal behavior of the hadronic phase 
formed in relativistic collisions~\cite{Andronic2012,Vovchenko2015,Rischke1991}. 

This article is organized as follows. 
Detailed descriptions of the formalism and methodology to calculate the thermodynamic properties are given in Sec. ~\ref{sec2}. 
In Sec. ~\ref{sec3}, we present the results along with a detailed discussion. Finally, we summarize our study in Sec. ~\ref{sec4}.
\section{Formalism}
\label{sec2}
\subsection{Hadron Resonance Gas with Excluded Volume}
The hadron resonance gas model describes the hadronic phase of the matter
by assuming that the system is composed of free hadrons of different species.  
It treats various hadrons and their resonances as point particles. However, the existence of repulsive interactions 
among the hadrons has already been known from the nucleon-nucleon scattering experiments. The excluded volume hadron 
resonance gas (EVHRG) model takes such repulsive interactions into account. 
The volume available for the 
hadrons to move is reduced by the volume that the hadrons occupy. 
The EVHRG model estimations for various thermodynamic quantities match 
the LQCD estimations up to $T \sim$ 140 MeV. It is also essential to take into account the effect of the Hagedorn mass 
spectrum, which includes the yet unmeasured higher masses of the hadrons, as they can have a significant contribution to 
the Equation of State (EoS) near the critical temperature ($T_{c}$).

In a Grand Canonical Ensemble (GCE) of ideal HRG (IHRG) formalism, the partition function of
the  particle $i$  can be written as~~\cite{Andronic2012};
\begin{equation}
\label{eq1}
ln Z^{id}_i = \pm \frac{Vg_i}{2\pi^2} \int_{0}^{\infty} p^2 dp\ ln\{1\pm \exp[-(E_i-\mu_i)/T]\}
\end{equation}
where $g_i$ and $E_i = \sqrt{p^2 + m_i^2}$ are the degeneracy and energy of the hadron 
$i$, respectively. The corresponding chemical potential is given by, 
\begin{equation}
\label{eq2}
\mu_i = B_i\mu_B + S_i\mu_S +Q_i\mu_Q,
\end{equation}
where $B_i$, $S_i$, and $Q_i$ are the baryon number, strangeness, and electric charge of the $i^{th}$ hadron. 
The $\mu_{B}$, $\mu_{S}$, $\mu_{Q}$ are the baryonic, strangeness and electric  chemical potentials 
respectively. 
The pressure $P_i$, energy density $\varepsilon_i$, number density $n_i$ and entropy density $s_i$ 
can be obtained from the partition function  by using the following relations: 
\begin{equation}
\label{eq3}
P^{id}_i(T,\mu_i) = \pm \frac{Tg_i}{2\pi^2} \int_{0}^{\infty} p^2 dp\ ln\{1\pm \exp[-(E_i-\mu_i)/T]\}
\end{equation}
\begin{equation}
\label{eq4}
\varepsilon^{id}_i(T,\mu_i) = \frac{g_i}{2\pi^2} \int_{0}^{\infty} \frac{E_i\ p^2 dp}{\exp[(E_i-\mu_i)/T]\pm1}
\end{equation}
\begin{equation}
\label{eq5}
n^{id}_i(T,\mu_i) = \frac{g_i}{2\pi^2} \int_{0}^{\infty} \frac{p^2 dp}{\exp[(E_i-\mu_i)/T]\pm1}
\end{equation}
\begin{align}
 s^{id}_i(T,\mu_i)=&\pm\frac{g_i}{2\pi^2} \int_{0}^{\infty} p^2 dp \Big[\ln\{1\pm  \exp[-(E_i-\mu_i)/T]\}\nonumber\\ 
&\pm \frac{(E_i-\mu_i)/T}{\exp[(E_i-\mu_i)/T]\pm 1}\Big]
 \label{eq6}
 \end{align}
 
The effect of higher states and resonances is included in IHRG by making use of the following parametrization for the Hagedorn mass spectrum ~\cite{Hagedorn1968};
\begin{equation}
\label{eq7}
\rho(m) = C\frac{\theta(m-M)}{(m^2+m_0^2)^a}\exp\Big(\frac{m}{T_H}\Big),
\end{equation}
where $M$ is the cut-off mass for selected hadrons and $T_H$ is the Hagedorn temperature chosen as 180 MeV for our analysis. For other parameters, we have chosen $C=0.05$ GeV$^{3/2}$, $m_0=0.5$ GeV and $a=5/4$ ~\cite{Vovchenko2015}. The inclusion of Hagedorn states leads us to adopt Boltzmann approximation. Thus, the previously defined thermodynamic properties are modified as;
\begin{align}
\label{eq8}
P^H(T,\mu) = &\frac{T}{2\pi^2} \int dm \int_{0}^{\infty} p^2 dp\ \exp\Big(-\frac{E-\mu}{T}\Big)\nonumber\\ & \times \Big[\sum_i g_i\delta(m-m_i)+  \rho(m) \Big]
\end{align}
\begin{align}
\label{eq9}
\varepsilon^H(T,\mu) = &\frac{1}{2\pi^2} \int dm\int_{0}^{\infty} p^2 E\ dp\ \exp\Big(-\frac{E-\mu}{T}\Big)\nonumber\\ &\times \Big[\sum_i g_i\delta(m-m_i)+  \rho(m) \Big]
\end{align}
\begin{align}
\label{eq10}
n^H(T,\mu) = &\frac{1}{2\pi^2} \int dm\int_{0}^{\infty} p^2 dp\ \exp\Big(-\frac{E-\mu}{T}\Big)\nonumber\\& \times \Big[\sum_i g_i\delta(m-m_i)+  \rho(m) \Big]
\end{align}
\begin{align}
 s^H(T,\mu) = &\frac{1}{2\pi^2} \int dm\int_{0}^{\infty} p^2 dp\ \exp\Big(-\frac{E-\mu}{T}\Big) \nonumber\\ &\times \Big(1+\frac{E-\mu}{T}\Big) \Big[\sum_i g_i\delta(m-m_i)+  \rho(m) \Big]
 \label{eq11}
 \end{align}
Repulsive interactions at short distances are introduced into IHRG by using thermodynamically consistent excluded volume approach~~\cite{Rischke1991}. This, in essence, is a van der Waals correction. The system volume $V$ is now replaced by the available volume $V_{av}$,
\begin{equation}
\label{eq12}
V\rightarrow V_{av} = V - \sum_{i}v_iN_i,
\end{equation}
where $v_{i} = 16\pi r_{i}^{3}/3$ is the excluded volume parameter 
(also called as eigen volume) \cite{Andronic2012,Vovchenko2015} and 
$N_{i}$ is the number of particles of type $i$.
From here on, we restrict ourselves to the case of equal volume parameter $v_i$ for all hadrons, $v = 16\pi r_h^3/3$. Use of Eq.~\ref{eq12} in grand canonical ensemble changes pressure through an iterative process ~\cite{Andronic2012,Vovchenko2015},
\begin{equation}
\label{eq13}
P^{ev}(T,\mu) = \kappa P^H(T,\mu),
\end{equation}
where, $\kappa$ is the excluded volume suppression factor given by,
\begin{equation}
\label{eq14}
\kappa = exp\bigg(-\frac{vP^{ev}}{T}\bigg).
\end{equation}
Accordingly, the thermodynamic entities like $\varepsilon$, $n$ and $s$ are  modified as follows; 
\begin{equation}
\label{eq15}
\varepsilon^{ev}(T,\mu) = \frac{\kappa\varepsilon^H(T,\mu)}{1+\kappa vn^H(T,\mu)}
\end{equation}
\begin{equation}
\label{eq16}
 n^{ev}(T,\mu) = \frac{\kappa n^H(T,\mu)}{1+\kappa vn^H(T,\mu)}
\end{equation}
\begin{equation}
\label{eq17}
 s^{ev}(T,\mu) = \frac{\kappa s^H(T,\mu)}{1+\kappa vn^H(T,\mu)}.
\end{equation}
\subsection{$Re$, $Kn$ and $Ma$ of the  Hadron Gas}
The transport properties of hadron gas can be investigated by using the Boltzmann Transport 
Equation (BTE) which is given by,
\begin{equation}
    \label{eq18}
    \frac{\partial f_{p}}{\partial t} + v^{i}_{p} \frac{\partial f_{p}}{\partial x^{i}} + F^{i}_{p} \frac{\partial f_{p}}{\partial p^{i}} = I(f_{p}),
\end{equation}
where $v^{i}_{p}$ is the velocity of the $i$th hadron and $F^{i}_{p}$ is the external force acting on it. $I(f_{p})$ is the collision integral which gives the change of distribution function due to collisions. The thermal equilibrium in a system is achieved and maintained by the 
collisions among the constituents. Under the 
relaxation time approximation the collision integral can be written as: 
\begin{equation}
    \label{eq19}
    I(f_{p}) = -\frac{(f_{p} - f_{p}^{0})}{\tau(E_{p})},
\end{equation}
where $\tau(E_{p})$ is the particle dependent relaxation time and $f_{p}^{0}$ is the equilibrium distribution function defined as;
\begin{equation}
    \label{eq20}
    f_{p}^{0} = \exp(-\frac{E_{i}-\mu}{T}).
\end{equation}
For a hadron gas at finite chemical potential, the shear viscosity can be obtained by 
using the following expression ~\cite{Kadam2015},
\begin{equation}
    \label{eq21}
    \eta = \sum_{i}\frac{1}{15T}\int \frac{d^3p}{(2\pi)^3}\frac{p^4}{E_i^2}(\tau_if^0_i + \Bar{\tau}_i\Bar{f}^0_i)
\end{equation}
where $\tau_i$ is the relaxation time of the hadron $i$ and $f^0_i$ is its corresponding 
equilibrium distribution function. The particle-dependent average relaxation time is given by,
\begin{equation}
    \label{eq22}
    \Tilde{\tau}^{-1}_i = \sum_{j}n_j\langle\sigma_{ij}v_{ij}\rangle,
\end{equation}
where $n_j$ is the number density of $j^{th}$ hadronic species,
$\sigma_{ij}$ is the cross section for the interaction between hadron $i$ and $j$, 
and $v_{ij}$ is the relative velocity between them. 
Furthermore, considering hadrons as hard spheres of equal radius $r_h$ (having constant cross section $\sigma = 4\pi r_h^2$), the thermal average of total cross section times relative velocity i.e. $\langle\sigma v\rangle$ can be calculated as ~\cite{Kadam2015,Cannoni2014,Gondolo1991},
\begin{equation}
\label{eq23}
\langle\sigma_{ij}v_{ij}\rangle=\frac{\int d^3p_id^3p_jv_{ij}\sigma{ij}f^0_if^0_j}{\int d^3p_id^3p_jf^0_if^0_j}=\sigma\langle v_{ij}\rangle.
\end{equation}
Now, rewriting the momentum integral in terms of energy of the hadrons and scattering angle $\theta$~\cite{Kadam2015,Tiwari2018} we get;
\begin{widetext}
\begin{equation}
    \label{eq27}
    \langle\sigma_{ij}v_{ij}\rangle = \frac{\sigma\int p_ip_jE_idE_iE_jdE_jd\cos\theta f^0_if^0_j\times \frac{\sqrt{(E_iE_j-p_ip_j\cos \theta)^2-(m_im_j)^2}}{E_iE_j-p_ip_j\cos\theta}}{\int p_ip_jE_idE_iE_jdE_jd\cos\theta f^0_if^0_j}
\end{equation}
\end{widetext}
where $E_i$ ($E_j$) is integrated in the limit $m_i$ ($m_j$) to $\infty$, respectively. 
The limit of integration of $\cos\theta$ runs from -1 to +1. Finally, the relaxation time is calculated using Eqs.~\ref{eq22} and ~\ref{eq27}. 

The Hagedorn states being highly unstable, decay rapidly and it is reasonable to assume that their presence affects the mean 
free paths of the hadrons in the medium. They also significantly affect the thermodynamics of hadronic matter close to the QCD critical temperature. Hence their contribution cannot be ignored. The contribution from these states to the shear viscosity can be estimated by using the approximation that the mean free path of such resonances is inversely related to their decay 
widths ~\cite{Kadamahep2019,Noronha2012,Noronha2009}. The decay width of Hagedorn states are
obtained by a linear fit to the decay widths of all resonances in the particle data book. Assuming that these particles have hard-core radius $r_h$, 
the excluded volume approximated contribution to shear viscosity from  Hagedorn states is given by,
\begin{eqnarray}
    \label{eq28}
    \eta_H =\frac{5T^{3/2}}{128\pi^{5/2}n^{H}(T)}\int m^{5/2} dm\ \rho(m) K_{\frac{5}{2}}\left(\frac{m}{T}\right).
\end{eqnarray}
where $n^{H}(T)$ is the number density of Hagedorn states in the medium and $K_n$ is the modified Bessel function of the second kind. 

The applicability of fluid dynamics in a system can be probed with the help of $Kn$. 
A small value for $Kn$ indicates a high degree of thermalization in the system due to 
frequent collisions between the constituent particles in the medium. 
The Knudsen number is defined by the ratio of characteristic microscopic to characteristic macroscopic lengths~\cite{Landau}, 
\begin{equation}
Kn=\frac{\text{microscopic length scale}}{\text{macroscopic length scale}}
\end{equation}
where the microscopic and the characteristic macroscopic length scales are taken as 
the mean free path ($\lambda$) and the linear dimension of the system ($D$) respectively. 
$D$ is the maximum possible value of the macroscopic length scale.  In central collisions of identical 
nuclei the maximum system size may be taken as the diameter of the nuclei 
however, the size will be smaller in peripheral collisions which depends 
on the value of the impact parameter.
For given $\lambda$, this will give the lower bound of $Kn$,  which will help in
finding out the domain of application of ideal hydrodynamics ($Kn\rightarrow 0$).
The first  order hydrodynamics $\sim Kn$ and the 
second order hydrodynamics $\sim Kn^2$, etc. 
We used the following expression for $Kn$ here,,
\begin{equation}
\label{eq30}
Kn = \lambda/D,
\end{equation}
where $\lambda = \frac{1}{\sqrt{2}n^{ev}\sigma}$. 
The inverse of Kundsen number is a measure of the average number of scattering 
that a hadron undergoes in the medium. 

The choice of characteristic macroscopic length scale as the linear size scale
can be justified for a system where the hydrodynamic variables do not change
appreciably for length scale smaller than the linear dimension of
the system {\it i.e.} when the gradient of hydrodynamic quantities are
small. Such a choice provides the lower bound of $Kn$ and upper bound of $Re$.
For small $Kn$ the ideal hydrodynamics is a good approximation 
for the description of the system as  the first and second order hydrodynamics
are proportional to $Kn$ and $Kn^2$ respectively.  

For a fluid with low $Kn$, the characteristics of the fluid and flow can 
now be studied by making use of $Re$ and $Ma$. 
The Reynold number is defined as~\cite{Landau}:
\begin{equation}
Re=\frac{\text{Inertial force}}{\text{Viscous force}}=\frac{l v\rho}{\eta}
\end{equation}
where $l$ is the characteristic macroscopic length scale $\sim$ spatial dimension over which
hydrodynamic variables do not change appreciably, $v$ is characteristic flow velocity, $\rho$ 
is the mass density of the fluid and $\eta$ is the shear viscosity.  For relativistic fluid  one 
replaces $\rho$  by $(\epsilon+P) =sT$, where $s$ is the 
entropy density (for non-zero baryonic chemical potential $\epsilon+P-\mu n =sT$). 
The Knudsen number is related with the inverse of Reynold number.
The characteristic macroscopic length scale $l$ appears through the shear force, $\sim \eta\partial v_i/\partial x_j\sim \eta v/l$ 
{\it i.e.} $l$ is to be estimated from the gradient of hydrodynamic quantities, for example, 
$l^{-1}\sim v^{-1}\partial v/\partial x$.
In the present work $l$ is chosen as the linear size ($D$) of the system {\it i.e.} $l=D$,  
which is the maximum possible value of $l$. This will give maximum value of $Re$ (keeping other parameters fixed)
and hence it imposes limit to the laminar flow. 

The Reynold number can also be  defined as the ratio of thermodynamic pressure ($P$) to bulk 
viscous pressure ($\Pi$) and thermodynamic pressure to the magnitude of shear stress tensor ($\pi^{\mu\nu}$) as~\cite{Denicol,Betz}:
$R_\Pi=(\epsilon+P)/\Pi$   and 
$R_\pi=(\epsilon+P)/\sqrt{\pi_{\mu\nu}\pi^{\mu\nu}}$ respectively which can be estimated realistically by solving
hydrodynamic equations for an expanding system formed in heavy ion collisions. 

Following simple ideal equation of state $\varepsilon = 3 P$, a limit on Reynold number  
can be imposed using causality~\cite{Chiu}. 
We estimate the $Re$ by using the expression given below,  
\begin{equation}
\label{eq29}
Re = \frac{D\langle v\rangle T}{\eta/s^{ev}},
\end{equation}
where $\langle v\rangle$ is the average of characteristic flow velocity as described below.
The expansion velocity of the fluid can be defined as~\cite{PhysKin}: 
\begin{equation}
\label{vav}
v_{\text{fluid}}(t,\vec{r})=
\int\frac{d^3p}{(2\pi)^3}\frac{\vec{p}}{E}f(E)/{\int\frac{d^3p}{(2\pi)^3}f(E)}
\end{equation}
In the non-relativistic domain, $\vec{p}/E$ may be replaced by $\vec{p}/m$.
The velocity of the particles (hadrons here) in the co-moving 
frame ($v^{\prime}$) is connected with the velocity in the laboratory frame ($v$)
through Lorentz transformation by the fluid velocity ($v_{\text{fluid}}$).
The average of $v^{\prime}$ is zero but not the average of $v$.
In eq.~\ref{vav} $f(E)$ is the phase space distribution  in the
co-moving frame which can be replaced
by local equilibrium distribution 
as a leading order approximation with space-time 
dependent temperature and chemical potential. For the purpose of estimating $Re$ 
one may take the 
space-time average of $v_{\text{fluid}}$ weighted by energy density of the fluid to
get $\langle v\rangle$. However, in the present work, the hydrodynamic
equations are not solved to get fluid velocity, therefore, we take this as
a parameter.  
It may be mentioned here that the average velocity of the fluid at rest may be non-zero
for non-central collision as indicated in Ref.~\cite{LPCsernai}.
We take different values of $\langle v\rangle=0.3c, 0.5c$ and $0.7c$ (see also~\cite{LPCsernai})
to understand the sensitivity of the results on the fluid velocity. The results are shown with $\langle v \rangle = 0.5 c$, which is in agreement with the values that can be estimated by using Eq.\ref{vav} for the chosen range of temperature.
The viscous correction to the distribution function has been neglected
as it is found to be small~\cite{vis_corr}. 
A high value of $Re$ is usually associated with the turbulence in the medium 
thereby creating eddies and through small eddies energy dissipation takes 
place in the fluid.
In the above two cases, we have used $D$ as the characteristic length scale.
This necessarily means that we have only two length scales which are:
the system dimension $D$ and mean free path $\lambda$. Therefore, in principle,
the length scale that appears in the expression for $Re$ must be more
than $\lambda$ and its upper limit should be $D$. 


Finally, the $Ma$ combines the fluid velocity and the speed of sound in the system and is written as,
\begin{equation}
\label{eq31}
    Ma = \langle v\rangle/c_s,
\end{equation}
where 
 \begin{equation}
\label{eq32}
    c_s^2 = \frac{\frac{\partial P}{\partial T} + \frac{\partial P}{\partial \mu_B}\frac{d\mu_B}{dT}}{\frac{\partial \varepsilon}{\partial T} + \frac{\partial \varepsilon}{\partial \mu_B}\frac{d\mu_B}{dT}},
\end{equation}
is the speed of sound in the system which reduces to $c_s^2 = (\frac{\partial P}{\partial \varepsilon})$ at vanishing baryon chemical potential. Here,
\begin{equation}
\label{eq33}
    \frac{d\mu_B}{dT} = \frac{s\frac{\partial n}{\partial T} - n\frac{\partial s}{\partial T}}{n\frac{\partial s}{\partial \mu_B} - s\frac{\partial n}{\partial \mu_B}}.
\end{equation}
The  $Re$, $Kn$ and $Ma$ obey the following relation \cite{Ollitrault2008}:
\begin{equation}
Re\times Kn = v\lambda\frac{\rho}{\eta}=\frac{v}{c_s}= Ma
\end{equation}
where we have used $\eta/\rho=\lambda c_s$ to obtain the above relation. 
This indicates the Mach number, $Ma$ is independent of length scale.

A small value of $Ma$ is usually associated with the incompressibility 
of the medium. 
It may be mentioned here that $Re$ and $Ma$ are independent
of each other as one of them depends on the macroscopic
dimension of the system. The $Ma$ would thus give similar results 
over different system sizes provided that the other available 
conditions are identical.

With the help of the formulae given above, we now move forward to estimate the dimensionless parameters $Re$, $Kn$ and $Ma$.
 
\section{Results and discussion}
\label{sec3}
We include all the hadrons and resonances up to a mass of 2.25 GeV listed in the particle data book~\cite{Zyla2020}
along with the Hagedorn spectrum. The thermodynamic properties are dependent on the eigen volume of hadrons which have to be chosen appropriately. It is dependent only on the unknown parameter $r_h$. The hadron-hadron interactions are not repulsive for an $r_h$ value comparable to the charge radii. Hence, it was argued that $r_h$ should be taken according to the hard-core radius known from nucleon-nucleon scattering. The repulsive interactions are mediated by $\omega$ mesons and the range of interaction is inversely proportional to the mass of the mediator. Based upon the above argument, we have chosen a hard-core radius, $r_h = 0.3$ fm, to incorporate the repulsive interaction. Also, there is not much deviation in charge radii between baryons and mesons, therefore, uniform hard-core radius ($r_h = 0.3$ fm) is considered for all hadrons~\cite{Munzinger1999,Kadam2015,Tiwari2018}.\\

The effect of system size on volume and particle numbers, considered infinite in the thermodynamic limit, can be implemented by providing a lower momentum cut-off to the integral over momentum space~\cite{Redlich2016,Bhattacharyya2015,Sarkar2017}. The finite-size effect is introduced by using a lower limit of momentum $p_{\text{cutoff}}$(MeV) = 197$\pi/D$(fm)~\cite{Sarkar2017}. Here $D = 2R$ is the characteristic system size, R being the system radius. We chose a few representative radii, $R =$ 1.5, 3, 5, 7 and 10 fm to cover system properties from high multiplicity pp to ultra-relativistic heavy-ion collisions based on experimentally obtained HBT radii~\cite{Sarkar2017,ALICE:2011kmy,ALICE:2015hav,ALICE:2019kno,Chatterjee2015}.
This choice of system size goes inline with the recent observations of finite hadronic phase lifetime observed in pp and heavy-ion collisions \cite{Sahu:2019tch,ALICE:2019xyr}.\\

\begin{figure}
\centering
\includegraphics[scale = 0.42]{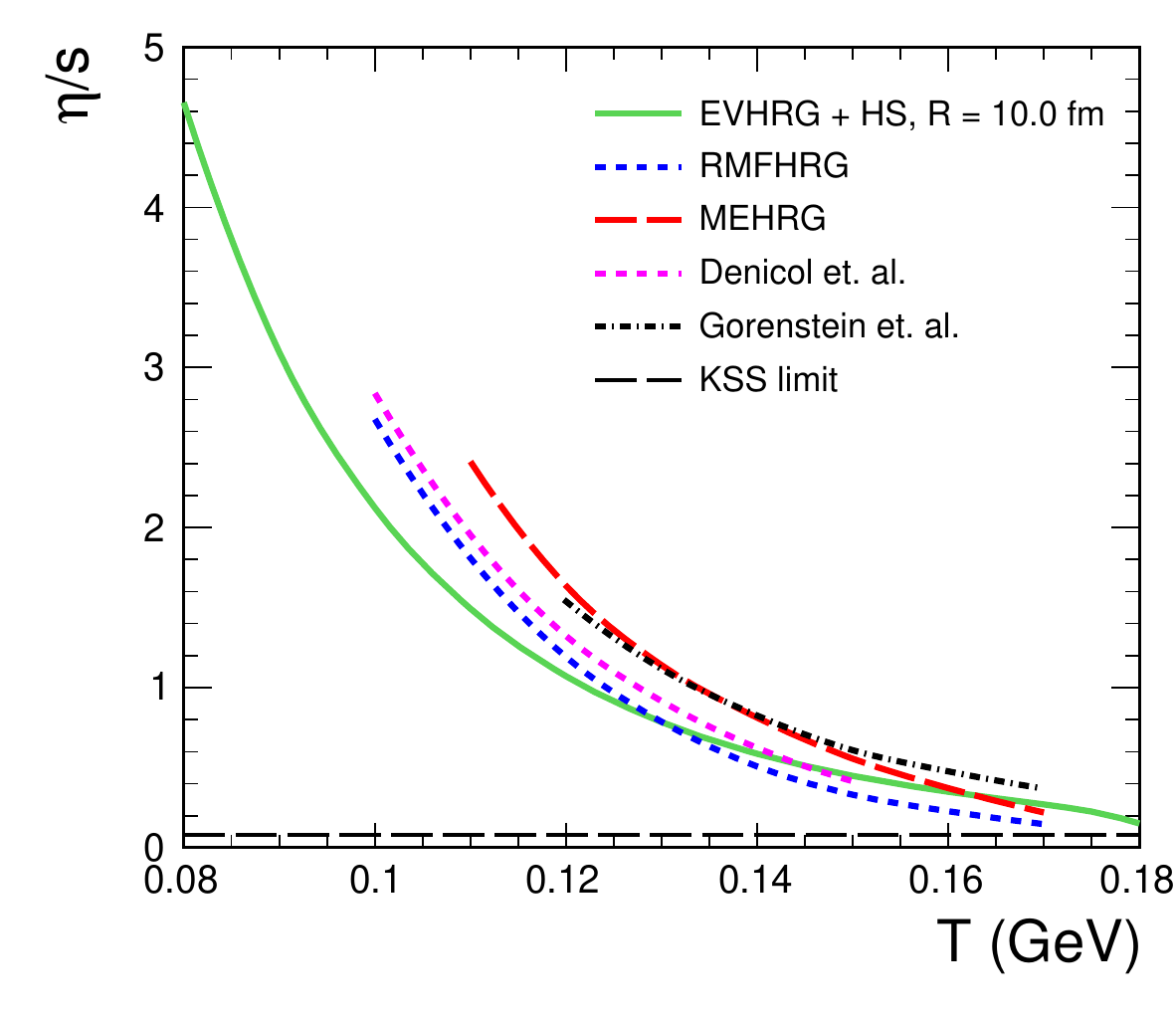}
\caption{Comparison of $\eta/s$ obtained for $R = 10.0$ fm with the estimations of other models at $\mu_{B} = 0.0$ GeV~\cite{Gorenstein2008,Denicol2013,Kadam2019,Kadamahep2019}.}
\label{fig1}
\end{figure}

To validate our study, we have compared our results with others in Fig.~\ref{fig1}. It shows the dependence of $\eta/s$ on system temperature for a fixed system size at zero chemical potential ($\mu_{B} = 0.0$ GeV). The AdS/CFT bound of
$\eta/s$ (also known as KSS limit), which is $1/4\pi$ ~\cite{Kovtun:2004de}, is also shown in the figure. It is 
observed that the $\eta/s$ in the given temperature range of the
hadronic phase shows a near-perfect fluid behavior towards higher temperatures. Our study is in good agreement with the values of $\eta/s$ obtained by Gorenstein et al., where the authors considered EVHRG within relativistic molecular kinetic theory~\cite{Gorenstein2008}. We have also compared our $\eta/s$ values with the results obtained from Chapman-Enskog theory, relativistic mean-field HRG  and EVHRG formalism with Hagedorn spectrum~\cite{Denicol2013,Kadam2019,Kadamahep2019}.\\

\begin{figure}[ht!]
\includegraphics[scale = 0.42]{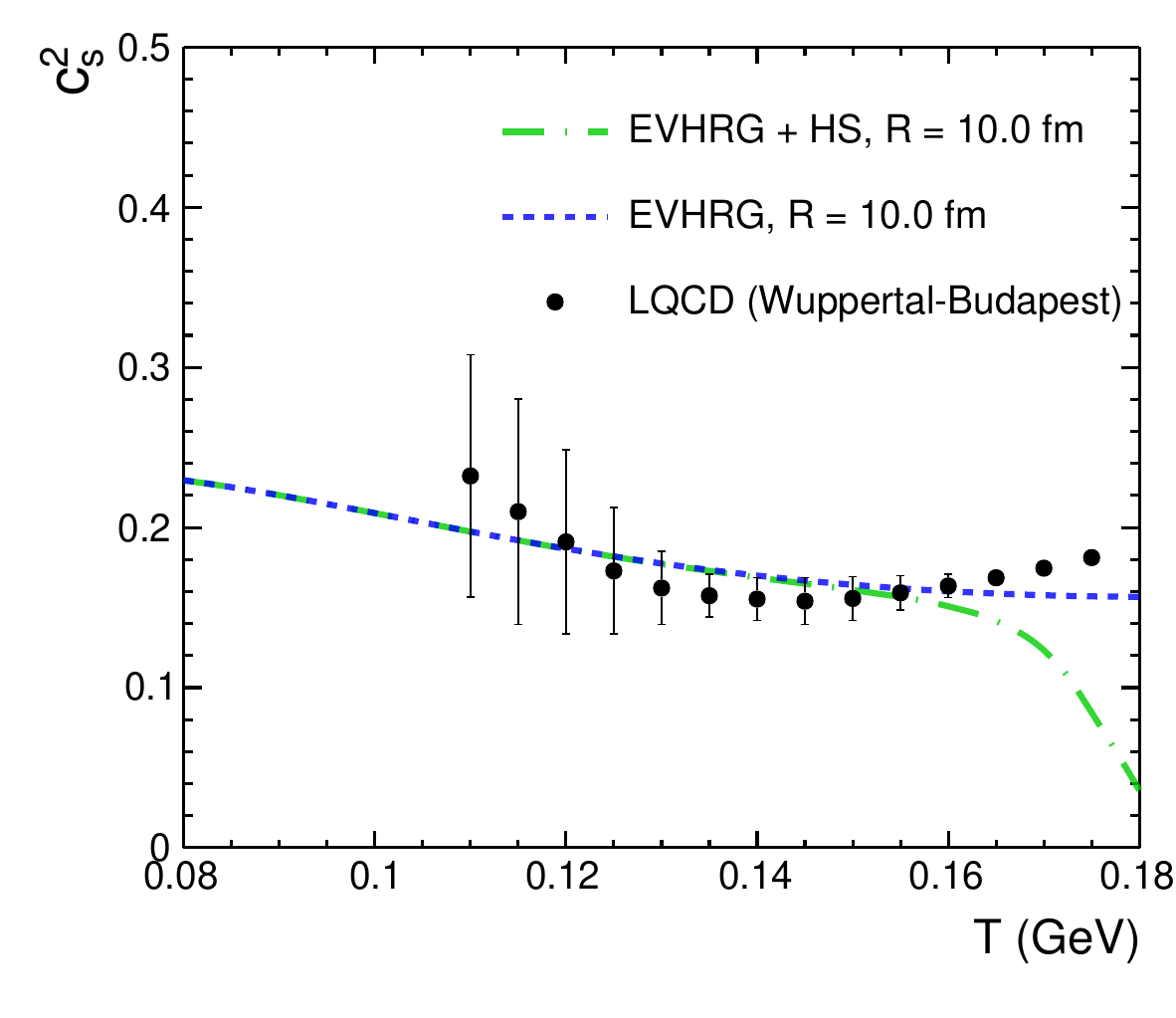}
\caption{The square of the speed of sound ($c_{s}^{2}$) as a function of temperature ($T$). This is compared with lattice data~\cite{Borsanyi2013}.}
\label{cs}
\end{figure}
Further, the dependence of the speed of sound on temperature is compared with lattice QCD  data at $\mu_{B} = 0.0$ GeV in Fig.~\ref{cs}. It is observed that  $c_s^2$  obtained in EVHRG formalism is in agreement with the LQCD data except at the high-temperature regime~\cite{Borsanyi2013},
where the HRG model is not a good approximation for the 
description of the system. The addition of the Hagedorn spectrum accentuates this deviation because of the rapid increase in energy density as compared to the pressure on approaching the Hagedorn temperature.\\

\begin{figure}
\includegraphics[scale = 0.42]{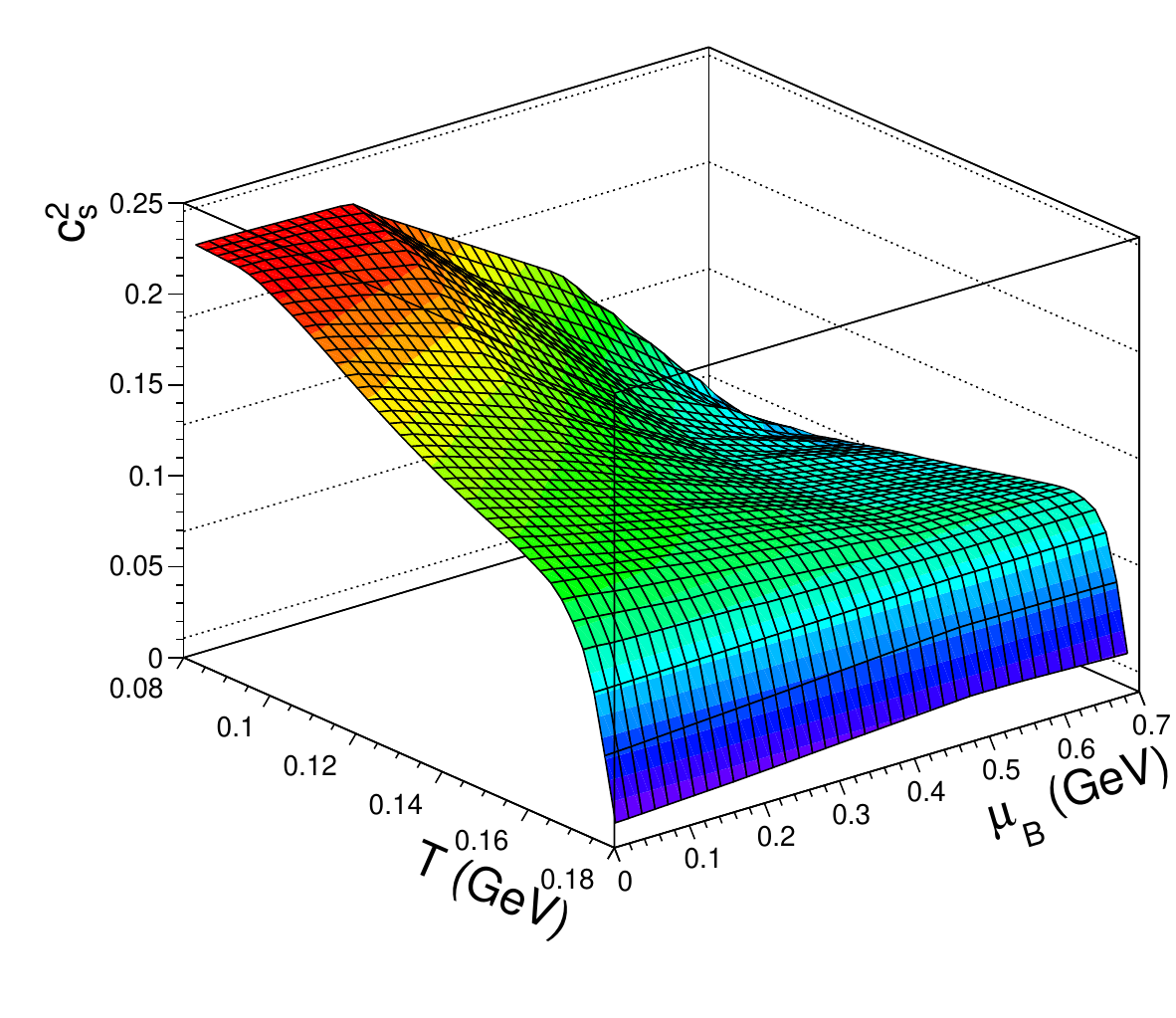}
\caption{The $c_{s}^{2}$ as a function of temperature and $\mu_{B}$ for $R = 10.0$ fm.}
\label{cs2}
\end{figure}

In Fig.~\ref{cs2} the variation of $c_{s}^{2}$ is  explored with temperature ($T$) and chemical potential ($\mu_{B}$). We have observed that increasing temperature at very low chemical potential ($\mu_{B} = 0-0.2$ GeV) gives similar results as discussed in Fig.~\ref{cs}. However, at $\mu_{B} > 0.2$ GeV with increasing $T$ 
speed of sound squared ($c_{s}^{2}$) decreases initially  up to $T\sim 0.140$ GeV 
and then start increasing with temperature. 
As $\mu_{B}$ increases the dip in $c_{s}^{2}$ shifts towards the lower $T$.
This smooth change in the behaviour of $c_{s}^{2}$ with high $\mu_{B}$ mimics the phase transition at $T<T_c \approx 0.160$ GeV. The sudden drop in the  $c_{s}^{2}$ at high $T$ is caused by the same effect of the Hagedorn states as seen in Fig.~\ref{cs}.\\

\begin{figure*}
\centering
\includegraphics[scale = 0.44]{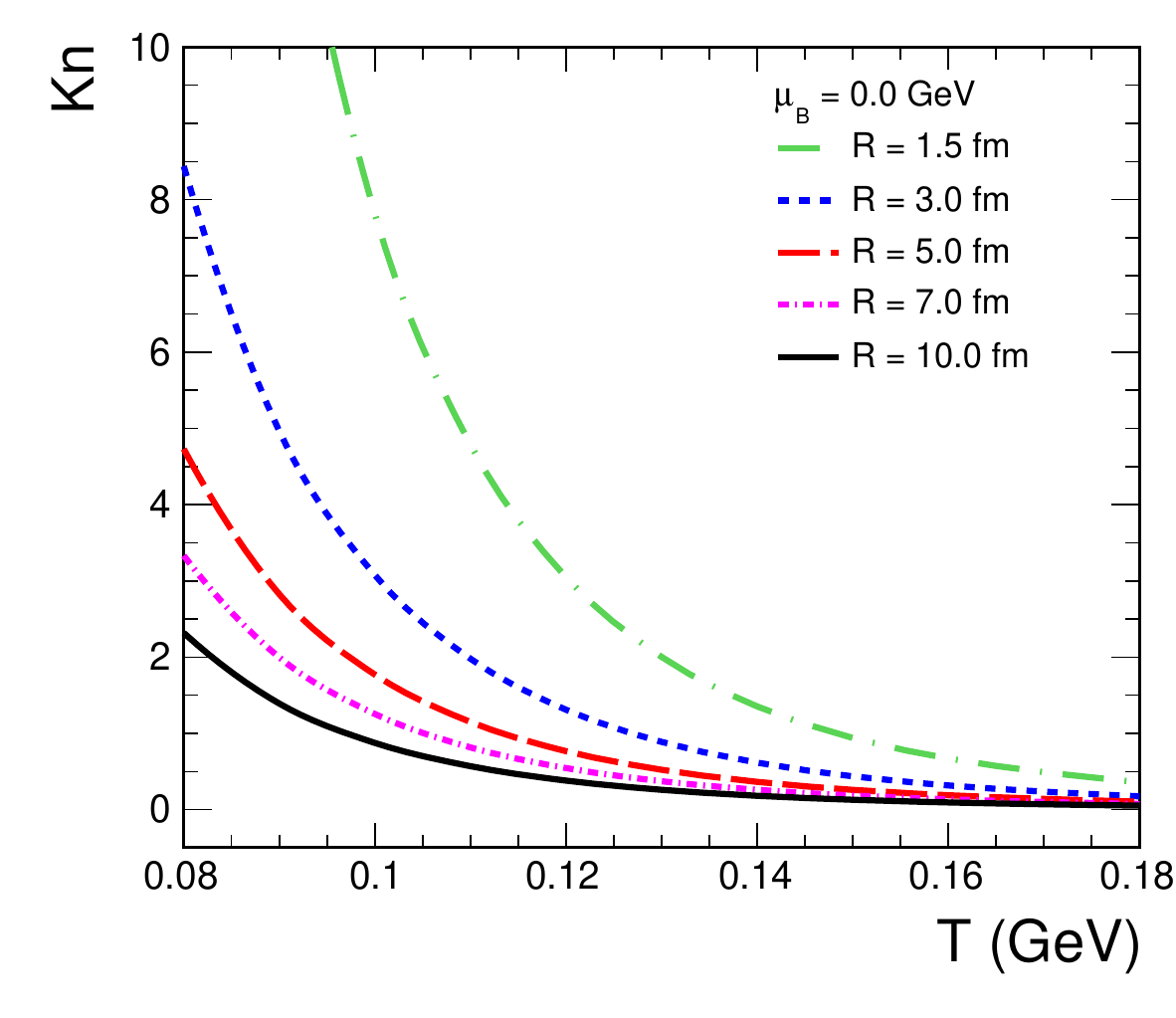}
\includegraphics[scale = 0.44]{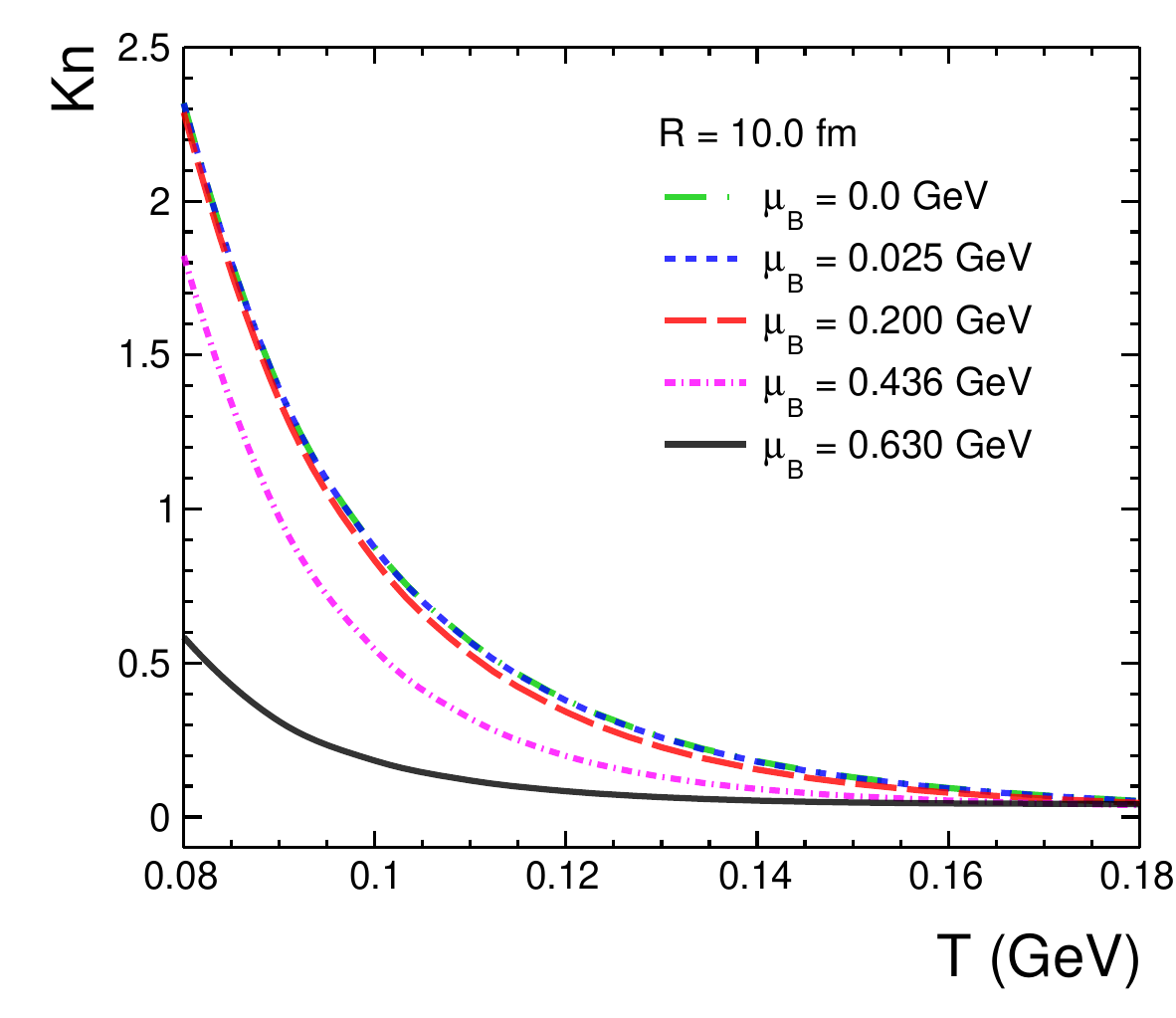}
\caption{Knudsen number, Kn, as a function of temperature for different values of system size, $R$ at $\mu_{B} = 0.0$ GeV (left panel) and for different values of $\mu_{B}$ at $R = 10.0$ fm, which is the case of larger systems (right panel).}
\label{fig3}
\end{figure*}

Large values of $Kn$, imply that the system is far from thermodynamic equilibrium, 
while small values (tending to zero), indicate a high degree of 
thermalization. 
Fig.~\ref{fig3} shows the dependence of $Kn$ on system radius. 
We observe a small value of $Kn$ at low temperatures for large $R$. This indicates that 
larger systems have a higher tendency to achieve thermodynamic equilibrium
as the hadrons undergo larger number of collisions.
The $Kn$ is small at high temperatures and does not vary much with system radius. This behavior is expected because at 
lower temperatures the  mean free path ($\lambda$) is larger, indicating less interaction among the hadrons. 
The variation of $Kn$ also indicates that the system formed at high temperatures 
is close to equilibrium (even for the lowest value of $R=1.5$ fm considered here) and therefore, 
we can apply hydrodynamics to study the evolution of the system. The variation of 
$Kn$ as a function of temperature for different 
chemical potentials is also shown in the right panel of Fig.~\ref{fig3}. 
We see that $Kn$ decreases with an increase in $\mu_{B}$.

\begin{figure*}[ht!]
\centering
\includegraphics[scale = 0.44]{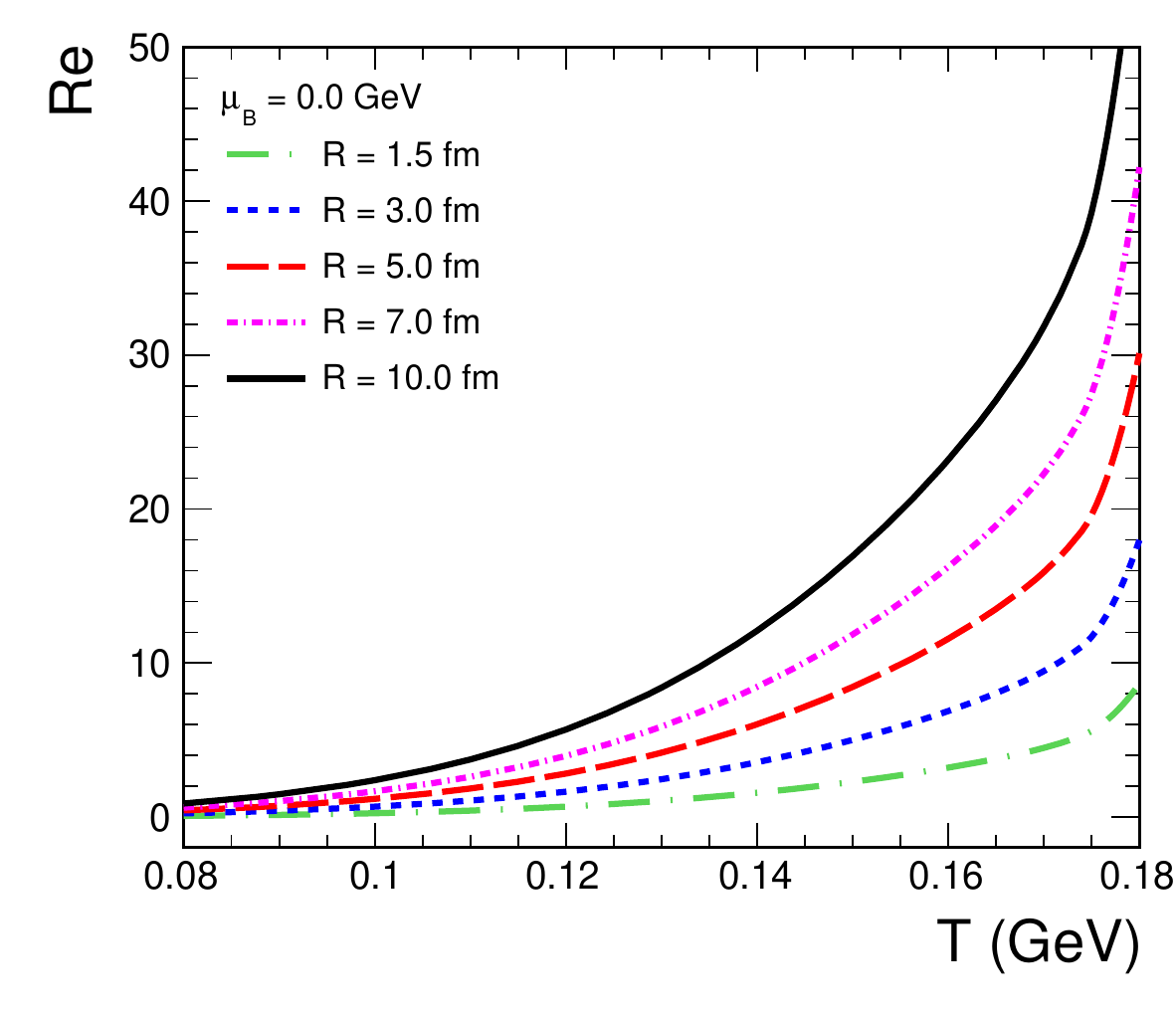}
\includegraphics[scale = 0.44]{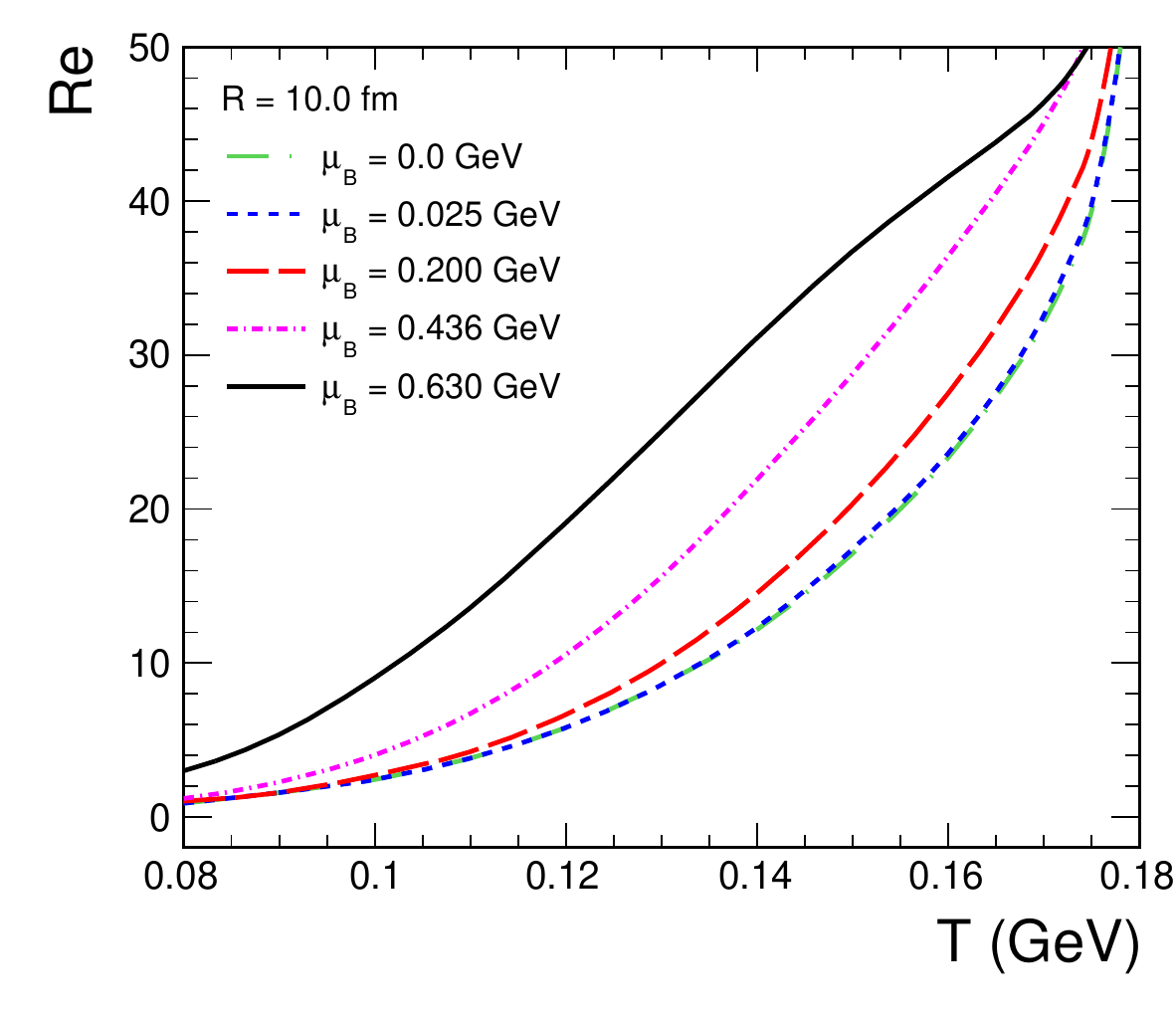}
\caption{Reynolds number as a function of temperature for different values of $R$ at $\mu_{B}$ = 0 GeV (left panel) and different values of $\mu_{B}$ at $R = 10.0$ fm, which is the case of heavy-ion collisions (right panel).}
\label{fig2}
\end{figure*}
$Re$, is an important parameter in hydrodynamics that contains information regarding instabilities 
in the fluid. A high value of $Re$ is usually associated with turbulence in the fluid. The inverse of Reynolds number, $Re^{-1}$, is generally seen as a proxy for $\eta/s$ in ultra-relativistic heavy-ion collisions as both of them are directly related. As a result of this relation, $Re>>1$ implies flow with low viscosity. We present the dependence of $Re$ 
on system size in Fig.~\ref{fig2}. We note that it increases with temperature for all values of $R$, indicating that the system becomes less viscous at higher temperatures and approaches a 'perfect fluid' limit. Fig.~\ref{fig2} also explores the variation of $Re$ with chemical 
potential (right panel). We have chosen five different values of chemical potentials; $\mu_{B} =$ 0.0, 0.025, 0.200, 0.436, and 0.630 GeV, which corresponds to LHC energies, RHIC at $\sqrt{s_{\rm NN}} = 200$ GeV and $\sqrt{s_{\rm NN}} = 19.6$ GeV, RHIC/FAIR at $\sqrt{s_{\rm NN}} = 7.7$ GeV and NICA at $\sqrt{s_{\rm NN}} = 3$ GeV respectively ~\cite{Tawfik2016,Munzinger2001,Cleymans2006,Khuntia2019}. The system radius is fixed at 10.0 fm, which roughly corresponds to ultra-relativistic heavy-ion collisions. At low temperatures, $Re$ increases with $\mu_{B}$ indicating that the viscosity of the system decreases with increasing $\mu_{B}$. As it appears, at 
higher temperature the  $Re$, becomes 
almost independent of the baryochemical potential of the system. 
The flow becomes turbulent for large $Re\gtrsim 2000$~\cite{McInnes2017}. The values of  $Re$
obtained here indicate that the flow is laminar for the system under consideration.
However, there may be other kinds of instabilities in the system for small $Re$, like
Kelvin-Helmholtz type ~\cite{LPCsernai}.\\

\begin{figure*}
\centering
\includegraphics[scale = 0.44]{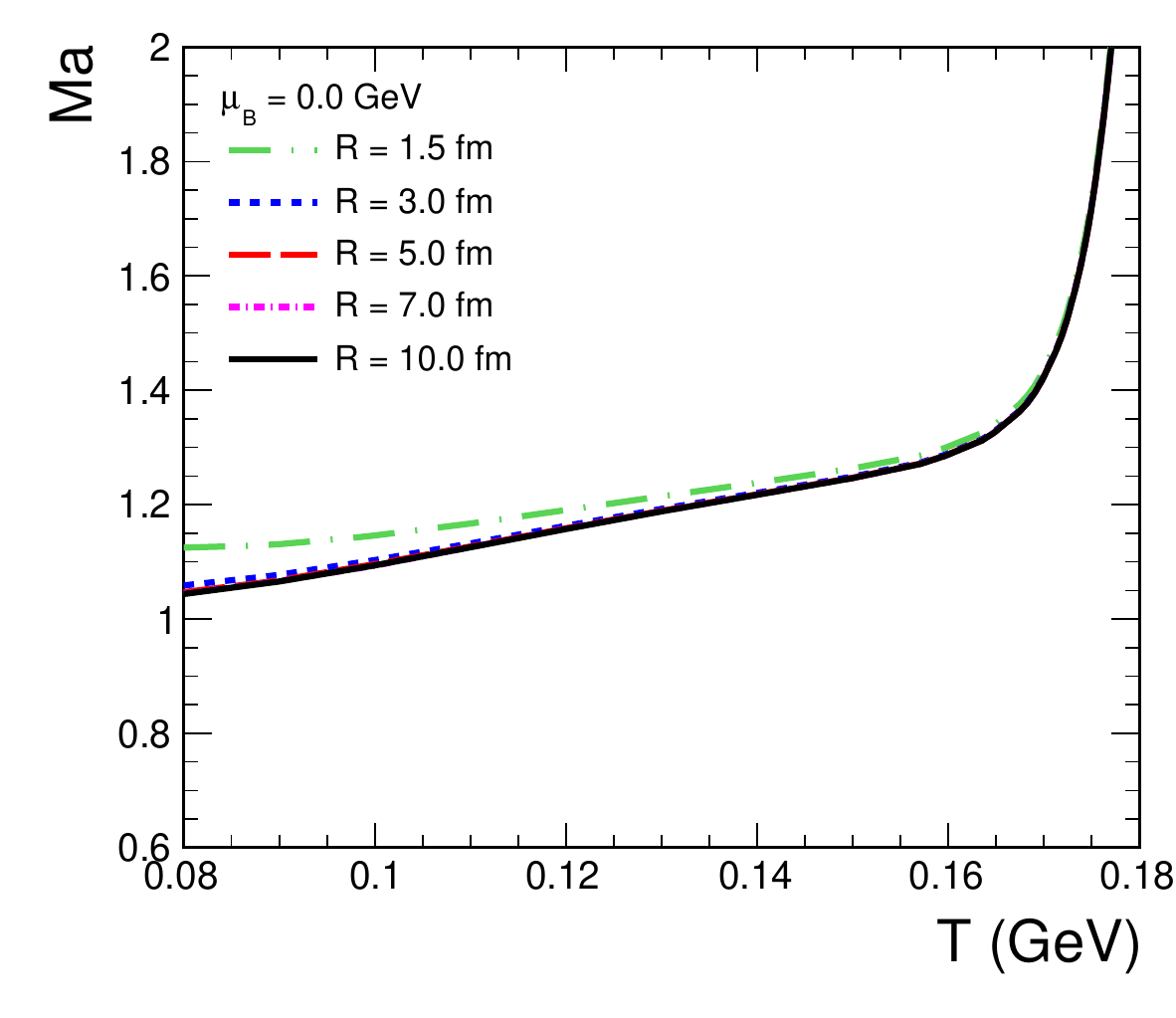}
\includegraphics[scale = 0.44]{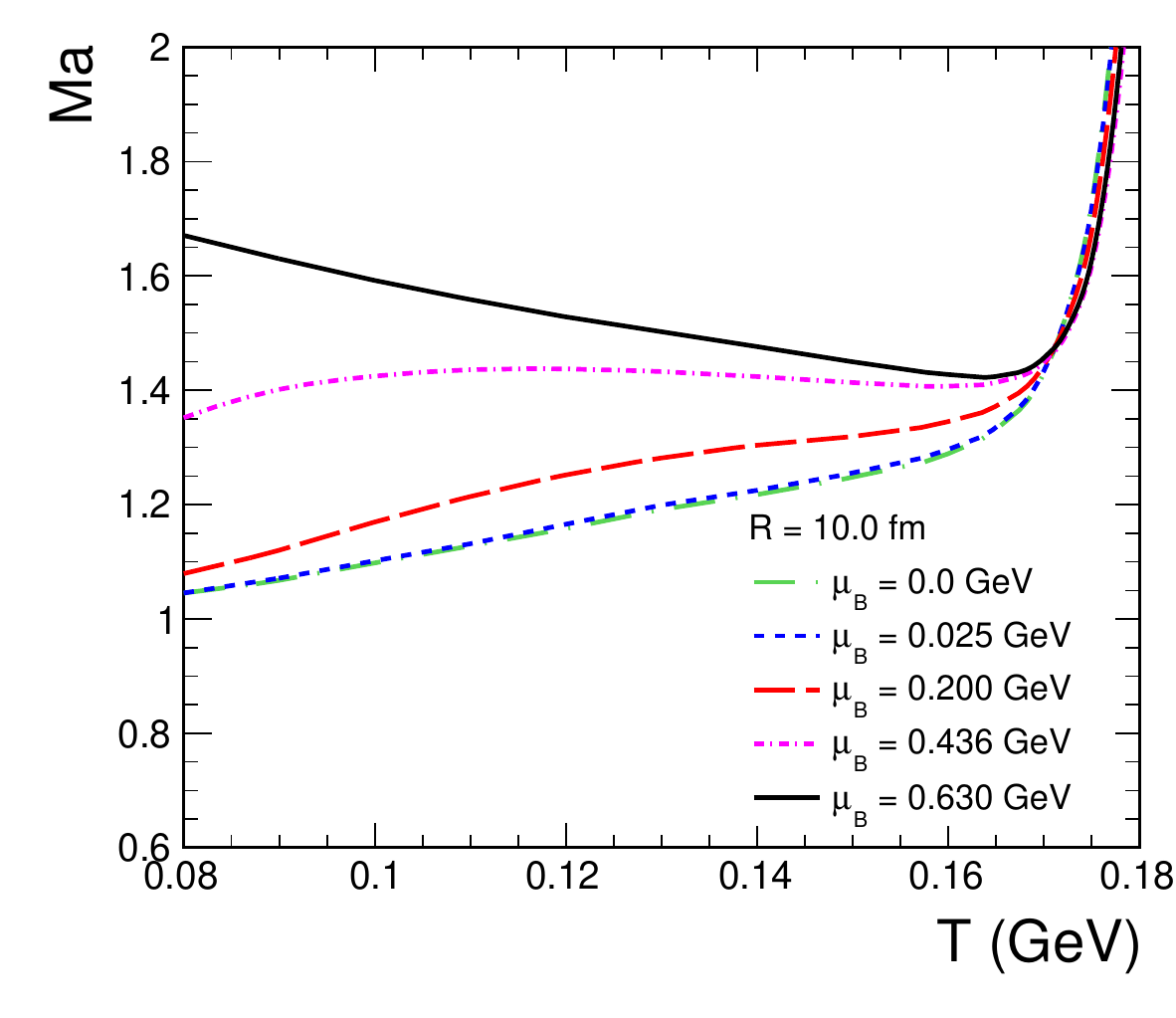}
\caption{Mach number, $Ma$, as a function of temperature for different values of system size, $R$ at $\mu_{B} = 0.0$ GeV (left panel) and for different values of $\mu_{B}$ at $R = 10.0$ fm, for larger systems (right panel).}
\label{fig4}
\end{figure*}
Supersonic flows significantly impact two-particle correlations because of the formation of Mach cones. 
The interest in Mach cones formed in ultra-relativistic heavy-ion collisions was sparked from the two-particle correlations observed at RHIC ~\cite{Aggarwal2010,Adare2010}. 
However, subsequent measurements at the LHC energies do not support these observations~\cite {Aad2014,Chatrchyan2014}. 
It is argued that the structure observed might be an artefact of incorrect background considerations ~\cite{Nattrass2016}. 
Although there is no direct experimental evidence for their existence, theoretical interest in Mach cones still 
persists as their disappearance could hint at a possible QCD critical point~\cite{Sarwar2021}. The dependence of $Ma$  
on $c_s$  makes it sensitive to the critical point (CP).
If $Ma << 1$, the density is almost uniform throughout the system and the fluid is incompressible indicating a very low bulk 
viscosity. Fig.~\ref{fig4} depicts the variation of $Ma$ with system radius. We observe that $Ma$ is almost independent 
of system radii beyond a particular temperature  
as expected with small deviations only due to the lower momentum cut-off. 
As temperature increases, it is observed that $Ma$ increases. 

It was observed that as $\eta/s$ increases, the typical Mach cone structure smears 
out and vanishes ~\cite{Bouras2014,Bouras2011}. Right panel of Fig.~\ref{fig4} 
displays the dependence of $Ma$ on $\mu_{B}$. We see that it increases with increasing $\mu_{B}$ and remains above the supersonic limit at all temperatures
considered here. This might be an indication of the early formation of a rapidly 
expanding system with shock waves at higher $\mu_{B}$ values. From Fig.~\ref{fig4}, 
it can also be inferred that since $Ma$ is greater than unity  
at all temperatures for all values of $\mu_{B}$ considered here, 
indicating that the system produced in ultra-relativistic collisions is 
compressible ~\cite{Ollitrault2008}. 

\begin{figure}
\centering
\includegraphics[scale = 0.44]{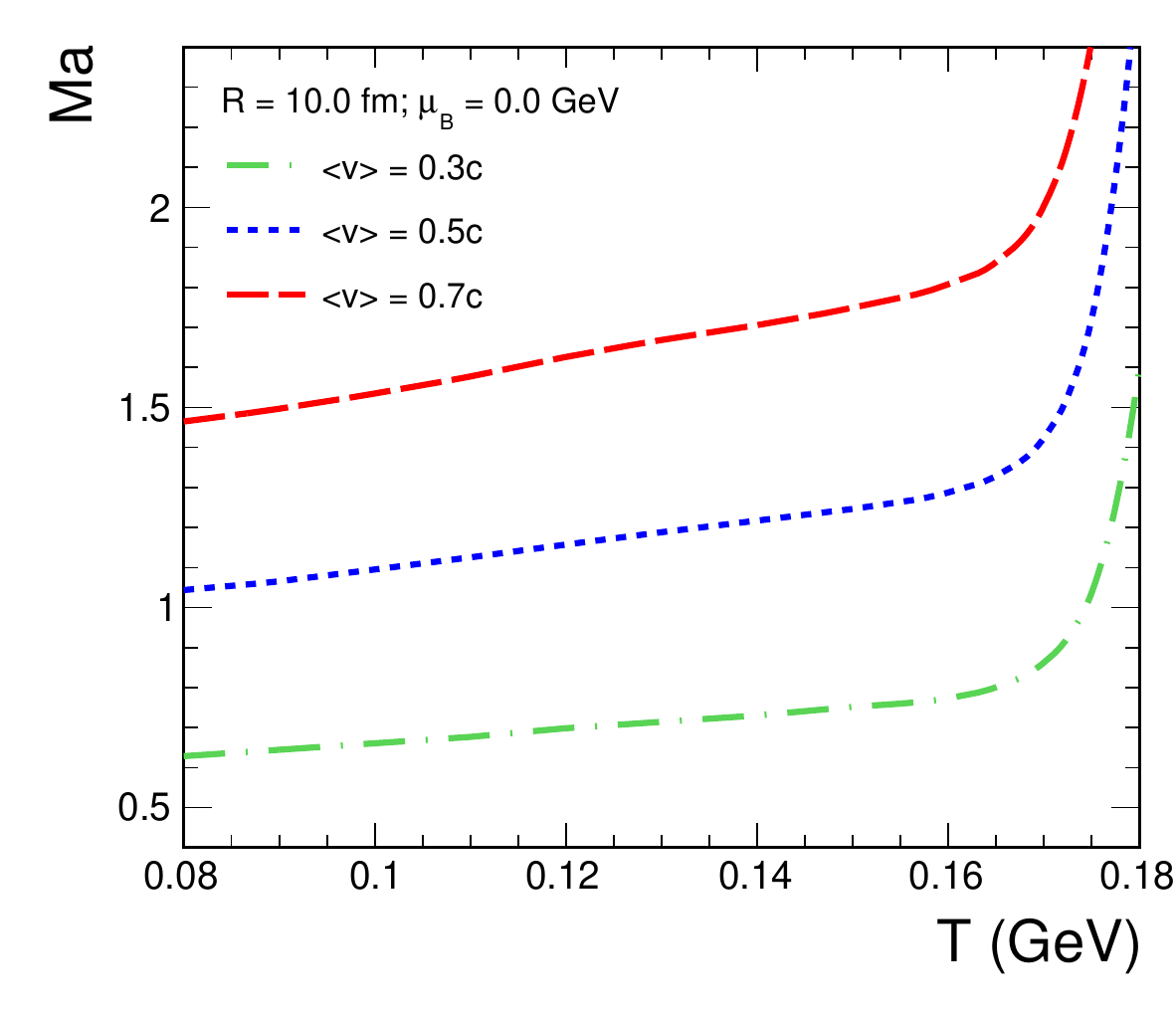}
\caption{The effect of variation of the fluid velocity parameter, $\langle v \rangle$, is shown. It is observed that a change in $\langle v \rangle$ results only in a scaling of the values without any change in the qualitative nature of the results.}
\label{fig5}
\end{figure}
Fig.\ref{fig5} depicts the effect of the magnitude of the
fluid velocity chosen here. It is observed that the qualitative dependence of $Ma$ on temperature 
remain unchanged. We can also deduce that the hadronic fluid considered is compressible for 
all values of fluid velocity greater than 0.5$c$. At lower velocities, the hadronic fluid is 
incompressible at low temperatures. A similar scaling can also be expected for $Re$ as 
$Re\propto \langle v \rangle$ as seen in Eq.\ref{eq29}.
\section{Summary}
\label{sec4}
In summary, we have estimated various relevant parameters like $Re$, $Kn$ and 
$Ma$ which assist
in characterizing  many-particle system. 
The dependence of these parameters 
on the system size, temperature, chemical  potential  for hadrons produced  in ultra-relativistic nuclear collisions
have been investigated. The obtained values ($Kn<<1$, $Ma \sim 1$ and $Re>>1$) indicate the occurrence of compressible low viscous flows at high temperatures 
close to the phase transition region ($T\sim 150-170$ MeV). The degree of 
thermalization of hadron gas estimated is comparable over different system sizes, 
hinting for the justification of the   application of hydrodynamics in 
interpreting the results from high multiplicity pp to heavy-ion collisions. 
This puts into question the applicability of pp collisions as a baseline for heavy-ion collisions at the high beam energies 
presently accessible at different particle accelerators.

Relativistic second-order causal viscous
hydrodynamics~\cite{Israel} is used extensively to analyze
the system formed in nuclear collisions at relativistic
energies, because the relativistic generalization
first order (Navier-Stokes) hydrodynamics is acausal
and it gives rise to unstable solutions.  
Application of second-order hydrodynamics needs shear and bulk viscous
coefficients, thermal conductivity, and various relaxation
and coupling coefficients. That is the second-order 
hydrodynamics introduces more unknown coefficients as inputs 
than its first-order version. Fluid dynamics can also be applied
to systems away from equilibrium by including higher
order gradients of hydrodynamic quantities~\cite{hydrobook}.
The inclusion of the order of these gradients will depend
on how far away the system remains from equilibrium.
The higher order gradients can be ignored for systems that 
are close to the equilibrium and hence a smaller number
of transport coefficients will suffice to characterize it.
We have estimated here the $Kn$ to examine the applicability
of the hydrodynamics, $Ma$ to determine
whether the fluid is compressible or incompressible. In the
relativistic domain, the fluid is compressible and hence
the bulk viscous coefficient will be required as input for hydrodynamic
simulation. The $Re$ decides whether the flow is
laminar or turbulent. For turbulent flow inclusion
of higher-order gradients will be crucial.
We have assumed that the hydrodynamic quantities change appreciably only over the linear dimension of the system. This assumption helps us to choose the characteristic macroscopic length scale as the linear dimension of the system for estimating $Kn$ and $Re$. Such a choice sets the lower and upper bound on $Kn$ and $Re$ respectively when other parameters (mean free path, average velocity, shear viscosity, etc.) are kept fixed. These bounds will be useful to figure out the applicability of ideal hydrodynamics. 
Therefore, the estimates
of $Kn$, $Re$, and $Ma$ from analysis of data will be useful
to understand fluid in a comprehensive manner.

\section*{Acknowledgment}
This research work has been carried out with financial support from DAE-BRNS, the Government of India,
Project No. 58/14/29/2019-BRNS of Raghunath Sahoo. For the research fellowship, Ronald Scaria acknowledges CSIR, Govt. of India. 
CRS and RS acknowledge the financial support under the above BRNS project. Further R.S. acknowledges
the financial support under the CERN Scientific Associateship, CERN, Geneva, Switzerland. The authors acknowledge the
Tier-3 computing facility in the experimental high-energy
physics laboratory of IIT Indore supported by the ALICE
project.

\end{document}